\documentclass[aps,prd,preprint,showpacs,showkeys,floatfix]{revtex4}
\usepackage{graphics}
\usepackage{bm}

\newcommand{\kti}[1]{\ensuremath{{{\bf k}_{#1}}_{\perp}}}
\newcommand{\kt}{{\textbf{k}_{\perp}}}
\newcommand{\lt}{{\textbf{l}_{\perp}}}
\newcommand{\lpt}{{\textbf{l}'_{\perp}}}
\newcommand{\smlpt}{{\textbf{\scriptsize l}'_{\perp}}}
\newcommand{\mt}{{\textbf{m}_{\perp}}}
\newcommand{\smmt}{{\textbf{\scriptsize m}_{\perp}}}
\newcommand{\qt}{{\textbf{q}_{\perp}}}
\newcommand{\EG}{{\textit{e.g.}}}
\newcommand{\IE}{{\textit{i.e.}}}
\newcommand{\EA}{{\textit{et al.}}}
\def\bra#1{\langle{#1}|}
\def\ket#1{|{#1}\rangle}

\begin{document}
\title{Separation of Soft and Hard Physics in Deeply Virtual Compton 
Scattering}
\date{\today}

\author{A. G{\aa}rdestig}
\email[Electronic address:]{agardest@indiana.edu}
\author{A.P. Szczepaniak}
\email[Electronic address:]{aszczepa@indiana.edu}
\author{J.T. Londergan}
\email[Electronic address:]{tlonderg@indiana.edu}
\affiliation{Indiana University Nuclear Theory Center, Bloomington, IN, U.S.A.}

\begin{abstract}
A model for deeply virtual Compton scattering, based on analytical 
light-cone hadron wave functions is presented and studied at energies 
currently accessible at Jefferson Laboratory, DESY, and beyond. 
It is shown that perpendicular vector components play an important role at 
$Q^2<10$~GeV$^2$ and that the meson-exchange diagrams are important at 
all energies. This could significantly impact the physical
interpretation of the underlying hadronic amplitudes. 
\end{abstract}


\maketitle

\section{Introduction}

Our knowledge about the partonic content of baryons has for several decades 
been obtained from experiments in the deep inelastic scattering (DIS) region.
At sufficiently high momentum transfer, the inclusive $ep\to e'X$ reaction is
very efficiently described in terms of parton distributions satisfying Bjorken 
scaling. The main reason for applicability of DIS to studies of hadron 
structure is that the process is factorizable into a hard part, 
calculable from perturbative quantum chromodynamics, and a soft 
part describing the quark-parton content of
 the hadron. This soft part gives the probability of finding a parton
 inside a proton with a certain fraction of the proton
 momentum.  The soft part is usually parametrized in terms of parton 
distribution functions.

\begin{figure}[ht]
\includegraphics{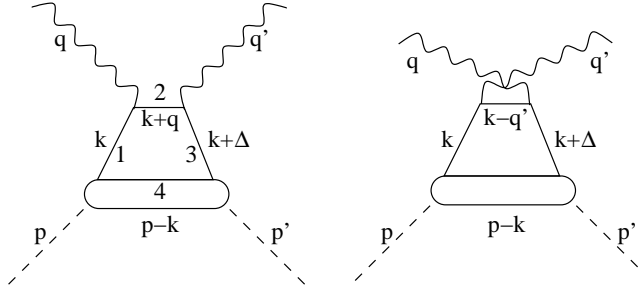}
\caption{The handbag diagrams, defining the kinematics and our notation.}
\label{fig:handbag}
\end{figure}

In recent years, exclusive reactions like $ep\to e'p'\gamma$ (deeply virtual 
Compton scattering or DVCS) or $ep\to e'p'M$ (hard electroproduction of 
mesons), have been explored experimentally and theoretically, since they 
promise to provide further insight into the parton distributions of hadrons. 
It turns out that even in certain exclusive reactions  
there is a factorization theorem that separates a hard, calculable  
part from a soft part~\cite{factor}. It is customary to define the soft 
interaction by introducing generalized parton distributions 
(GPD's)~\cite{ji,vdH,RadI,KrollDVCS}, that describe the transition amplitude 
for removing a parton with a certain momentum fraction and then putting it 
back with a different momentum fraction.  The GPD's contain information 
about correlations between different parts of the proton wave function, or, 
equivalently, between partons at different locations 
inside the proton. This geometrical view has been explored by Burkardt, 
Diehl, Ralston and Pire, and others~\cite{Burkardt}. The calculations of 
DVCS have recently been extended to include twist-three effects~\cite{tt}.
The factorization of DVCS is conveniently represented by the handbag 
diagrams (Fig.~\ref{fig:handbag}), where the lower blobs are parametrized 
in terms of generalized parton distributions. The same diagrams have also 
been applied to wide angle Compton scattering and photon annihilation into 
hadron--anti-hadron pairs~\cite{KrollOCS}.

In the Bjorken limit ($Q^2\to\infty$, $x_B=Q^2/2p\!\cdot\! q$ constant), there 
are four independent GPD's; $H(x,\zeta,t), E(x,\zeta,t), 
\widetilde{H}(x,\zeta,t)$, and $\widetilde{E}(x,\zeta,t)$, where $x$ and 
$\zeta$ are respectively the light-cone momentum fractions of the struck 
quark and real photon, while $t=\Delta^2$ is the 
momentum transfer squared. Three physically different regions can be 
distinguished for $x$ and $\zeta$.  The domain $0<\zeta<x<1$ 
($\zeta-1<x<0$) corresponds to the removal and return of a quark 
(antiquark) with momentum fractions $x$ ($\zeta-x$)
and $x-\zeta$ ($-x$), respectively. In the remaining region $0<x<\zeta$, 
the photon scatters on a virtual quark-antiquark pair, extracted from the 
proton.  The latter situation could be described as a meson exchange, but 
is usually incorporated into the general parametrization of the 
GPD's~\cite{vdH,tt}.  An alternative interpretation is to use light-cone 
perturbation theory~\cite{LB} and regard the meson exchange as a higher 
Fock state of the proton splitting of a quark--anti-quark pair~\cite{BDH}.
With this notation, $\zeta\to x_B$ (Bjorken $x$) in the limit 
$Q^2\to\infty$ ($\Delta$ fixed).  
In the limit of forward scattering (DIS), the $H$'s 
reduce to the quark density and quark helicity distributions, \EG
\ $H(x,0,0) = q(x)$ and $\widetilde{H}(x,0,0) = \Delta q(x)$.
The $E$'s do not appear in DIS.  They are unique to the off-forward 
exclusive processes and provide information not accessible through other 
means.  Finally the $x$-integrated GPD's are related to the nucleon form
factors~\cite{ji}.  

An alternative approach to DVCS using light-cone quark wave 
functions~\cite{LB} was suggested in Refs.~\cite{Krolllcwf,BDH}. 
The authors have used this idea to calculate DVCS on an electron in QED 
for large $Q^2$, assuming that the electron temporarily splits into a 
virtual electron-meson pair.

There are ambitious efforts under way to measure DVCS (and the hard 
exclusive meson photo-production) at a number of facilities, 
in particular at DESY~\cite{HERMES} and JLab~\cite{CLAS}. Experiments using 
high-energy muon beams are planned for the COMPASS facility at 
CERN~\cite{COMPASS}.  At JLab 
energies, the competing Bremsstrahlung or Bethe-Heitler (BH)~\cite{BH} 
process is larger than DVCS. However, by carrying out interference 
measurements 
($e^+/e^-$ beam charge asymmetry and various spin asymmetries), the
BH amplitude cancels out and only a BH$\times$DVCS interference remains. 
At DESY the energy is sufficiently large that DVCS becomes larger than 
BH, though 
the measurements have low statistics, insufficient to extract  
differential cross sections.  However, both facilities are able to 
measure asymmetries for relatively low beam energies and momentum transfers. 
In this paper we will investigate to what extent the `leading' 
amplitude actually dominates in this kinematic region. In order to 
address this question, we employ a model using effective analytic 
quark-diquark wave functions.  This model allows 
us to study the dynamics of the amplitudes excited at these low energies. 
Some preliminary results were reported in Ref.~\cite{GSL}.
This model is similar to the one presented in Ref.~\cite{BDH}, but we 
calculate DVCS on a proton and focus especially on the features at low 
$Q^2$, and we keep higher-twist terms.

\section{Formalism}

In this paper, we treat the proton as a quark-diquark state.  For 
simplicity, we drop the proton spin degrees of freedom. 
Furthermore we will only consider the valence quark flavors, \IE, we 
ignore any sea quark or higher Fock state contributions to the proton 
wave function.  Thus, for a proton, the $SU(3)$ operator $\lambda/\sqrt{2}$ 
is replaced by the $SU(2)$ (isospin) operator $\tau/\sqrt{2}$. 
In future articles we will relax these constraints on our model, but 
this simplified model will suffice for the present purpose of studying the 
$Q^2$ dependence and analytic structure of the DVCS amplitudes.
Our model proton state has the form 
\begin{eqnarray}
     \ket{p} & = & \int\frac{dx_1d^2\kti{1}}{16\pi^3}
        \frac{dx_2d^2\kti{2}}{16\pi^3}
    \frac{16\pi^3\delta(x-x_1-x_2)\delta^2(\kti{1}+\kti{2})}{\sqrt{x_1x_2}}
        \nonumber \\ &&
        \frac{\delta_{{\lambda_1\lambda_2}}}{\sqrt{2}}
        \frac{\tau^j_{f_1f_2}}{\sqrt{2}}\frac{I_{c_1c_2}}{\sqrt{3}}
      \phi(x_i,\kti{i})b^{\dagger}(x_1,\kti{1})d^{\dagger}(x_2,\kti{2})\ket{0},
\end{eqnarray}
where $\lambda_i$, $f_i$, and $c_i$ are the quark spin (helicity), flavor 
(isospin), and color indices.  The proton state normalization is 
\begin{equation}
        \langle p'|p\rangle = 16\pi^3\delta(x'-x)
        \delta^2({\bf p}'_{\perp}-{\bf p}_{\perp}).
\end{equation}

\begin{figure}[ht]
\includegraphics{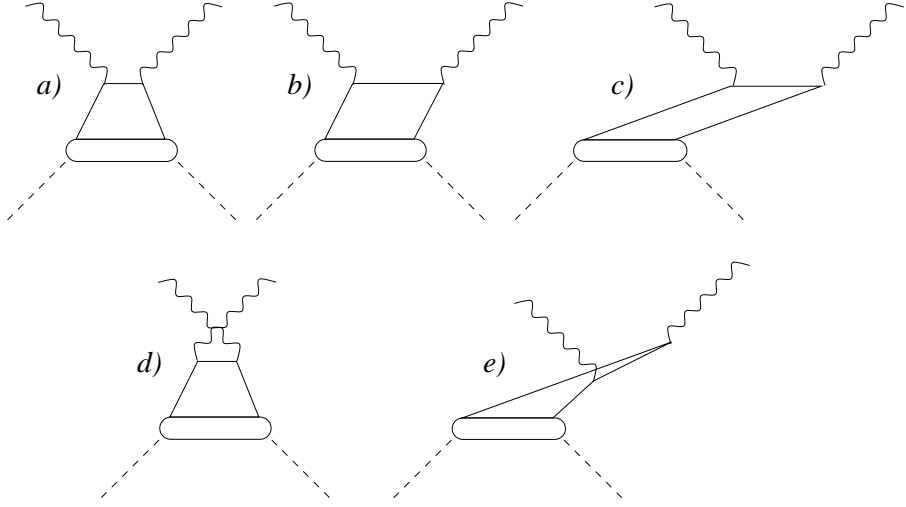}   
\caption{The five different time-ordered `handbag' diagrams.}
\label{Fig:diag}
\end{figure}

We start from the Fourier transform of the $\gamma^{\ast}p\to\gamma p'$ 
amplitude (in light-cone coordinates)
\begin{equation}
       T^{++} = -i\int d^4y e^{iq'\cdot y}\bra{p'}TJ^{+}(y)J^{+}(0)\ket{p},
\end{equation}
where $J^{+}(y)=\bar\psi(y)\gamma^+\gamma_5\frac{\tau}{\sqrt{2}}\psi(y)$ is 
the electromagnetic current, and $p(p')$ and $q(q')$ are respectively 
the four-momenta of the initial (final) hadron and photon.  
In light-cone time-ordered perturbation theory this expression can be 
expanded to give the five different one-loop diagrams (assuming light-cone 
gauge $q^+=0$) shown in Fig.~\ref{Fig:diag}.
 
These diagrams are essentially the same as those used in Brodsky  
\EA~\cite{BDH} for DVCS on an electron.  The two approaches differ in 
that Brodsky \EA\ include the electron spin and treat the system as a 
fermion-vector composite, while we assume the diquark to be a fermion 
and approximate the proton as a scalar.  In addition, we employ 
soft Gaussian form factors for our proton wave functions.  The time-ordered 
diagrams of Fig.~\ref{Fig:diag} are represented by the integrals
\begin{eqnarray}
T^{++}_a & = & 2(p^+)^2\int_{\zeta<x<1} \frac{dxd^2\kt}{16\pi^3}
        \frac{\phi^{\dagger}(z,\lt)\phi(x,\kt)}{
        \left[M^2+\left(\frac{1}{x_B}-1\right)Q^2-\frac{m^2+{\smmt}_a^2}{x}
        -\frac{m_R^2+{\smmt}_a^2}{1-x}+i\epsilon\right]}, 
\label{eq:TDefa} \\
T^{++}_b & = & 2(p^+)^2\int_{0<x<\zeta} \frac{dxd^2\kt}{16\pi^3}
     \frac{\left(m_R^2-\frac{m^2+\smlpt^2}{z'}-\frac{M^2+\smlpt^2}{1-z'}\right)
        \varphi(z',\lpt)\phi(x,\kt)}
        {\left(-\frac{\zeta(m^2+{\smmt}_b^2)}{x(\zeta-x)}\right)} 
	\nonumber \\ 
        &\times& \left[ M^2+\left(\frac{1}{x_B}-1\right)Q^2-
        \frac{m^2+{\smmt}_a^2}{x}
        -\frac{m_R^2+{\smmt}_a^2}{1-x}+i\epsilon \right]^{-1}, \\
        T^{++}_c & = & 2(p^+)^2\int_{0<x<\zeta} \frac{dxd^2\kt}{16\pi^3}
     \frac{\left(m_R^2-\frac{m^2+\smlpt^2}{z'}-\frac{M^2+\smlpt^2}{1-z'}\right)
                        \varphi(z',\lpt)\phi(x,\kt)}
        {\left(-\frac{\zeta(m^2+{\smmt}_b^2)}{x(\zeta-x)}\right)
        \left(\frac{\Delta^2}{\zeta}-\frac{\zeta(m^2+{\smmt}_c^2)}{x(\zeta-x)}
                \right)}, \\
        T^{++}_d & = & 2(p^+)^2\int_{\zeta<x<1} \frac{dxd^2\kt}{16\pi^3}
        \frac{\phi^{\dagger}(z,\lt)\phi(x,\kt)}
        {\left(\frac{1}{1-\zeta}\left(M^2-\Delta^2-\frac{Q^2}{x_B}\right)-
                \frac{(1-\zeta){\smmt}_d^2}{(x-\zeta)(1-x)}-\frac{m^2}{x-\zeta}
                -\frac{m_R^2}{1-x}\right)}, 
\label{eq:TDefd} \\
        T^{++}_e & = & 2(p^+)^2\int_{0<x<\zeta} \frac{dxd^2\kt}{16\pi^3}
     \frac{\left(m_R^2-\frac{m^2+\smlpt^2}{z'}-\frac{M^2+\smlpt^2}{1-z'}\right)
                        \varphi(z',\lpt)\phi(x,\kt)}
        {\left(-\frac{\zeta(m^2+{\smmt}_e^2)}{x(\zeta-x)}\right)
        \left(\frac{\Delta^2}{\zeta}-\frac{\zeta(m^2+{\smmt}_c^2)}{x(\zeta-x)}
                \right)}.
\label{eq:TDefe}
\end{eqnarray}
In Eqs.\ \ref{eq:TDefa}-\ref{eq:TDefe} we define the kinematic 
variables $z=(x-\zeta)/(1-\zeta), z'=(\zeta-x)/(1-x)$, and the relative 
momenta are defined as 
\begin{eqnarray}
        \lt & = & \kt+(1-z)\Delta_{\perp} \nonumber \\
        \lpt & = & -(1-z')\kt-\Delta_{\perp} \nonumber \\
        {\mt}_a & = & \kt+(1-x)\qt \nonumber \\
        {\mt}_b & = & \kt+\qt-\frac{x}{\zeta}\qt' \nonumber \\
        {\mt}_c & = & \kt+\frac{x}{\zeta}\Delta_{\perp} \nonumber \\
        {\mt}_d & = & \kt-(1-z)\qt' \nonumber \\
        {\mt}_e & = & -\kt-\frac{x}{\zeta}\qt'.
\end{eqnarray}
Here $x=k^+/p^+>0$ and $\zeta=q'^+/p^+$ are the longitudinal momentum
fractions of the struck quark and the real photon, the quantities 
$m, m_R$, and $M$ are respectively the quark, remnant (diquark), and hadron 
masses, and $\Delta = q-q'$. 

In the limit $Q^2\to\infty$, the denominators of Eqs.~(\ref{eq:TDefa}) and 
(\ref{eq:TDefd}) are proportional to $x-\zeta+i\epsilon$ and 
$x+\zeta-i\epsilon$.  These are the leading twist expressions of Ji~\cite{ji}. 
In our model these leading-twist diagrams ($a$ and $d$ of 
Fig.~\ref{Fig:diag}) are integrated over $\zeta<x<1$ for which the 
Compton scattering occurs on a quark extracted from the proton. In 
principle, scattering on an anti-quark would have the same 
form, but it would require a different (higher Fock state) wave function.
For the skew diagrams $b, c$, and $e$ of Fig.~\ref{Fig:diag}, the 
photon scatters on a quark-antiquark pair split off from the proton.
The proton wave function is represented by the analytic form
\begin{equation}
        \phi(x,\kt) = N\exp\left[-\frac{1}{\beta^2}
     \left(\frac{m^2}{x}+\frac{m_R^2}{1-x}+\frac{\kt^2}{x(1-x)}\right)\right],
\label{eq:wfdef}
\end{equation}
where $\beta=0.6911$~GeV is chosen such that 
\begin{equation}
  F(q^2)=\int\frac{dxd^2\kt}{16\pi^3}\phi(x,(1-x)\qt+\kt)\phi(x,\kt)
\end{equation}
agrees with the dipole form factor~\cite{dipole} for $Q^2<1$~GeV$^2$. 
The skew diagrams require knowledge of the wave function $\phi(z',\lpt)$ for 
the diquark splitting into a hadron and quark (the lower right-hand corner 
of diagrams $b$, $c$, and $e$ of Fig.~\ref{Fig:diag}). 
The form of this wave function will eventually be restricted by exclusive 
data, but in this paper is arbitrarily chosen to be of the form of 
Eq.~(\ref{eq:wfdef}), with $m_R$ and $x$ replaced by $M$ and $z'$.
Since the relative phase of the two wave functions is not known, an 
arbitrary complex phase factor $\exp(i\phi)$ is introduced between the 
regular and skew diagrams.

The full $T$ matrix could be related to the purely kinematic $T^{++}$ 
component by the Lorentz structure 
$T^{\mu\nu}\propto\tilde{p}^{\mu}\tilde{n}^{\nu}
+\tilde{n}^{\mu}\tilde{p}^{\nu}-g^{\mu\nu}$, derived by Ji~\cite{ji}. 
In this expression $\tilde{p}$ and $\tilde{n}$ are two lightlike 
four-vectors that project out the $+$ and $-$ components of Ji's formalism, 
and have been Lorentz transformed into our reference system. 
The Bethe-Heitler cross section is calculated from standard methods, using 
the dipole parametrization~\cite{dipole} of both the electric and magnetic 
proton form factors. Since we do not include the proton spin, the 
BH$\times$DVCS interferences are not calculated.

\section{Results}
\begin{figure}[ht]
\includegraphics{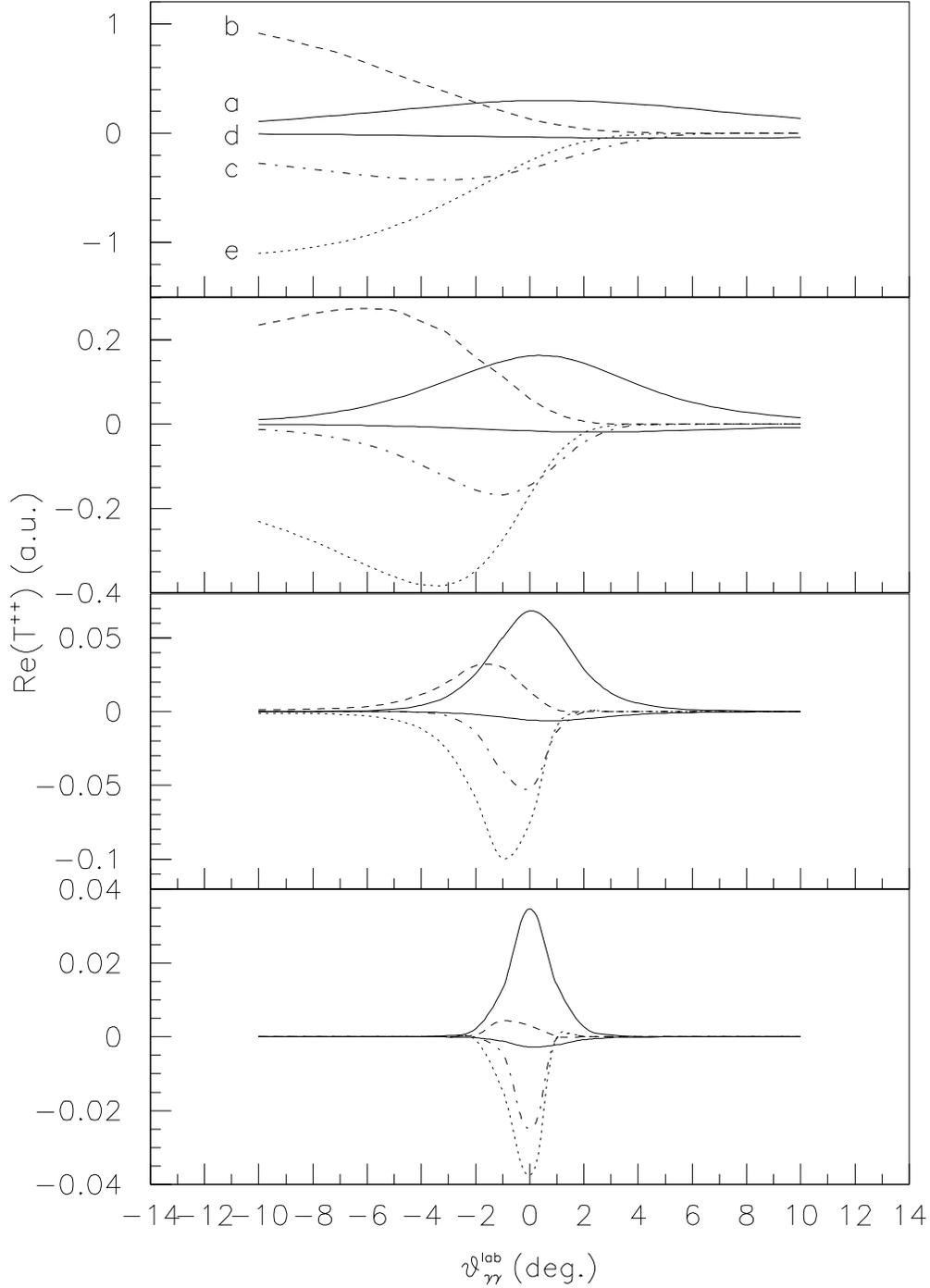}   
\caption{The ${\Re\rm e}T^{++}$ matrix elements as functions
of the laboratory angle between the virtual and the real photon for in-plane
scattering. The labels correspond to the ones of Fig.~\protect\ref{Fig:diag}
(the lower solid line always corresponds to diagram $d$).
The calculations are for $x_B=0.35$ and $Q^2=2,4,10,20$~GeV$^2$ 
from top to bottom, using an arbitrarily assigned phase angle 
$\phi=0^{\circ}$.}
\label{Fig:Tpp}
\end{figure}

The $T^{++}$ matrix elements have been calculated for present JLab and DESY
kinematics, using standard Monte Carlo techniques. 
These calculations incorporate both the principal value (${\Re\rm e}T^{++}$)
and $\delta$-function (${\Im\rm m}T^{++}$) parts.
The angular distribution and
$Q^2$ evolution of the real parts are plotted in Fig.~\ref{Fig:Tpp}.
The laboratory angle $\theta_{\gamma\gamma'}$ between the virtual and real 
photons is defined for in-plane kinematics such that it is positive for 
$\phi=0$ and negative for $\phi=180^{\circ}$, where $\phi$ is the 
azimuth angle between the final electron and the real photon, with 
${\bf\hat{q}}$ as the polar axis. Note that this angle convention is the 
reverse of the one in Ref.~\cite{GSL}.

The leading twist diagram ($a$ of Fig.~\ref{Fig:diag}) is the largest 
for most scattering angles, with diagrams $c$ and $e$ giving 
contributions of roughly the same order.  In our calculations, the 
contribution from the crossed diagram $d$ is generally related to 
the leading-twist term by $d\sim-a/10$.  This situation is quite different 
from DIS, where in general $d\ll a$.  This difference can be understood 
by realizing that DVCS is an exclusive process with an on-shell 
off-forward final photon 
($\Rightarrow{\Im\rm m}T_d=0$, ${\Re\rm e}T_d \neq0$), while DIS has an 
off-shell photon (${\Im\rm m}T_d \neq0$), so that the imaginary part of the 
cut diagram is the only contribution to the DIS cross section.  
Since we are interested in the real part of diagram $d$ (which is absent 
in DIS), and DIS uses the (negligible) imaginary part (which is absent in 
DVCS), the two reactions turn out to be quite different and do not 
measure the same physics.

In the physical region of the DVCS process, diagrams $a$ and $b$ possess 
a cut on the real axis, while the other diagrams never get vanishing 
denominators for a massless final photon.  Thus the imaginary part of the 
DVCS amplitude, which arises from the $\delta$-function part of the 
propagators, has contributions from the first two diagrams ($a$ and $b$) 
only.  These imaginary parts are not plotted in Fig.~\ref{Fig:Tpp} but 
are included in the calculation of the cross section.

Because of its two hard propagators, diagram $b$ is suppressed 
in the high $Q^2$ limit~\cite{BDH}, and in our calculation  
this suppression does becomes significant, but only for large  
$Q^2>10$~GeV$^2$, \IE, well above present JLab energies. 
On the other hand, diagrams $c$ and $e$ remain important at all energies and 
could be interpreted as contributing to meson exchange mechanisms, since the
quark-antiquark pair could form an intermediate (off-shell) meson. 
The skew diagrams $b, c$, and $e$ yield non-symmetric angular 
distributions, reflecting the shifted denominators in these cases. 

The difference in analytic structure also explains why diagram $d$ 
gives a smaller contribution than diagram $a$, since the $Q^2$ terms add 
up in the denominator of $d$, while they have opposite signs for $a$. This 
feature is closely related to the behavior of the two leading-twist 
propagators of Ji~\cite{ji}.  These two terms have the form  
$1/(x-\xi+i\epsilon)$ and 
$1/(x+\xi-i\epsilon)$, where $x$ and $\xi$ are momentum fractions related to 
$\frac{1}{2}(p^++p'^+)$ instead of $p^+$. In this notation there is a
cancellation for $x=\xi$ (scattering on quark) or $x=-\xi$ (antiquark).
The important difference between the two approaches is that Ji neglects
four-vector components, \EG, $\Delta_{\perp}$ that do not give large 
scalars in the Bjorken limit, while we explicitly keep them since we are 
interested in the behavior at low and moderate $Q^2$.

\begin{figure}[ht]
\includegraphics{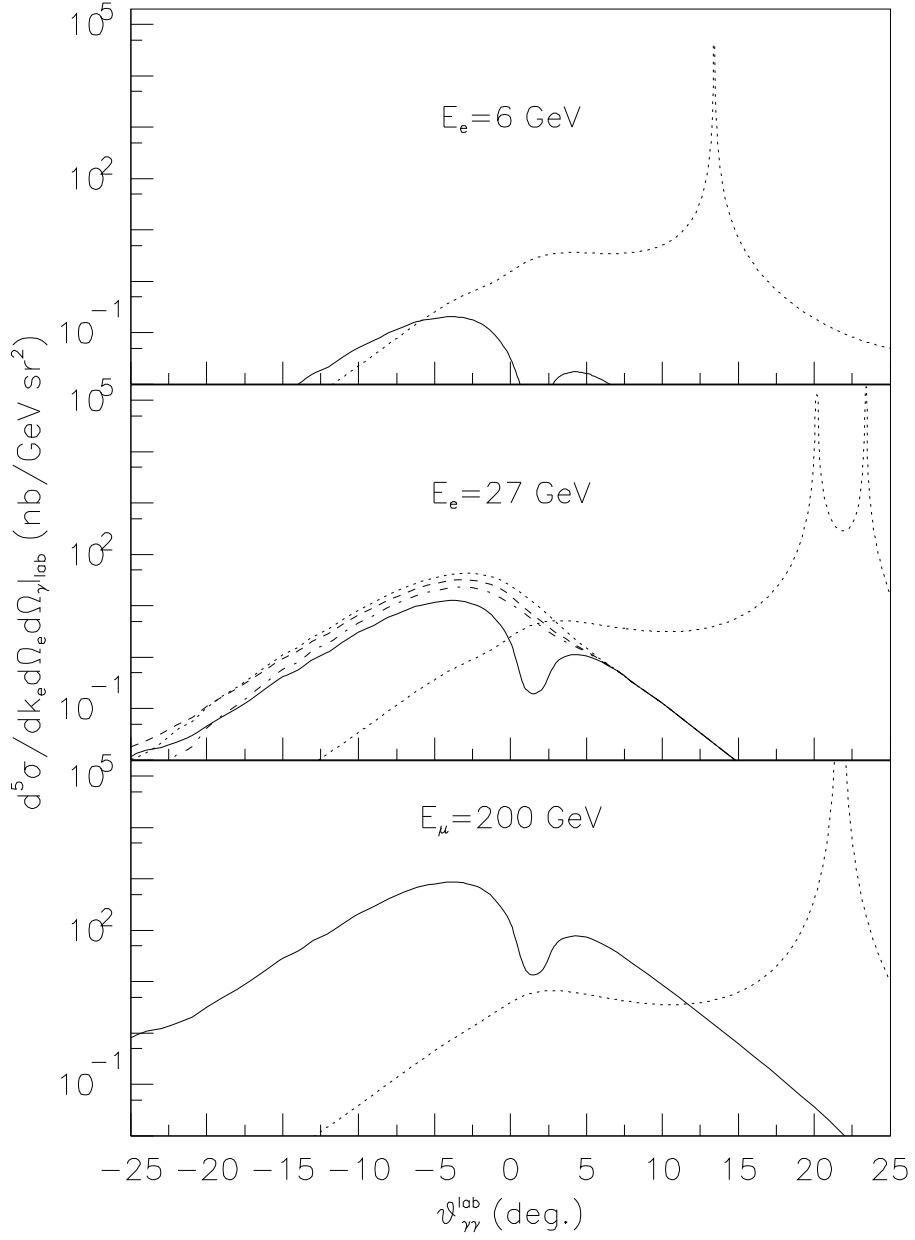}
\caption{Differential cross section for DVCS (solid line) and BH 
(right-most dotted line) at $x_B=0.3$, $Q^2=2$~GeV$^2$, and beam energies of 
the JLab, DESY, and CERN experiments. 
In the second graph the phase between diagrams $a,d$ and $b,c,e$ is 
$0^{\circ}$ (solid line), $90^{\circ}$ (dashed line), $180^{\circ}$ 
(dotted line), and $270^{\circ}$ (dot-dashed line).} 
\label{Fig:QDep}
\end{figure}

In Fig.~\ref{Fig:QDep} our predictions for the differential cross 
sections for DVCS and BH are shown for a range of beam energies, and for 
various choices of the relative phase between the regular and skew 
diagrams.  These plots are comparable to those published in \cite{vdH} 
for the kinematics corresponding to the JLab~\cite{CLAS}, 
DESY~\cite{HERMES}, and CERN~\cite{COMPASS} experiments. 
Only for very large $Q^2$ is it possible to separate DVCS from BH when 
only unpolarized cross sections are available.  
The dip in the DVCS cross section at $\approx2^{\circ}$ for zero phase 
angle reflects the fact that the real parts add up to zero there. 
If such a structure would be observed (at large beam energies), it could
help constrain the phase between the regular and skew diagrams.

\begin{figure}[ht]
\includegraphics{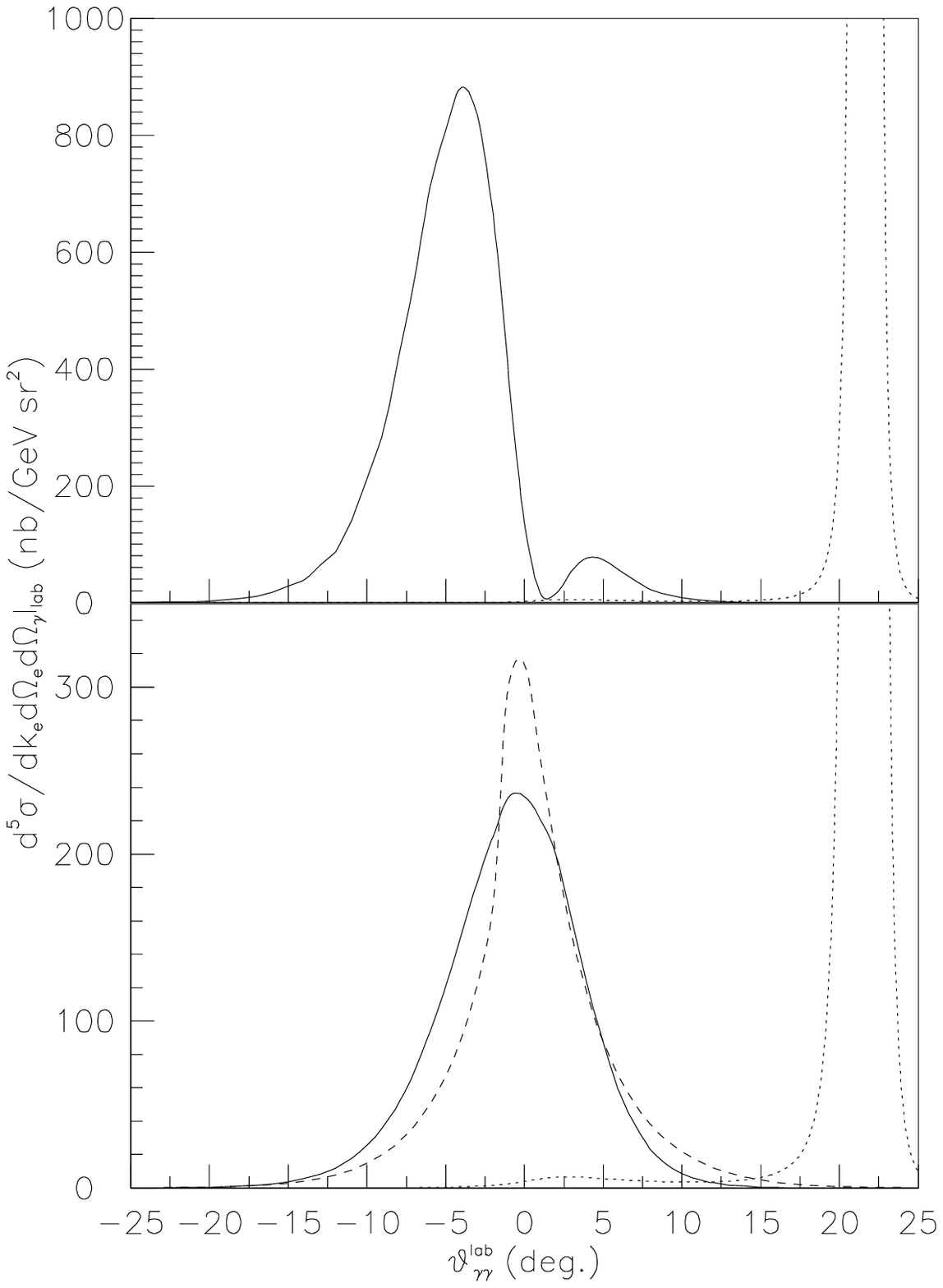}
\caption{Effect of excluding the skew diagrams, for energy 
$E_\mu = 200$ GeV.  Upper graph: skew diagrams included.  Lower graph, 
only the leading twist diagrams $a$ and $d$ are included. Lower 
graph, dashed line: only 
leading twist diagrams, and setting  
$\kt=0$ in the denominators (dashed line). Note that 
the two plots use different linear scales.}
\label{Fig:skew}
\end{figure}

The effect of the skew diagrams on the cross section is seen 
Fig.~\ref{Fig:skew}.  When they are removed (lower graph, solid line), the
dip in the cross section disappears, but the cross section for backward 
angles also decreases by roughly an order of magnitude. 
However, the cross section is still asymmetric.
If in the remaining leading twist diagrams $a$ and $d$ we set $\kt=0$ 
in the energy denominators (lower graph, dashed line), we end up with a 
narrower, almost symmetric 
distribution. At small angles this shift increases the cross 
section by 30--40\% and for backward angles it decreases it by a factor of 
about two.

The skewed behavior of the angular distribution is hence caused by a 
combination of two effects; the backward shifted meson exchange (skew) 
diagrams, and the presence of perpendicular components in the propagators.

\section{Conclusions and outlook}
We have introduced a model for DVCS, using simple analytic quark wave 
functions and retaining all components of every four-vector in the 
scattering process.  This model enables us 
to investigate the five single-loop diagrams and their 
relative importance for DVCS at JLab and HERA energies.
Our results indicate that at present JLab energies ($Q^2<4$~GeV$^2$), all 
of the diagrams of Fig.~\ref{Fig:diag} (except possibly the crossed 
diagram $d$) contribute to the DVCS cross sections.   
In particular it is necessary to include diagram $b$, despite its two hard 
propagators, since the suppression of this amplitude is significant only 
for very large $Q^2$.  Even at kinematics appropriate for the upgraded 
(12~GeV) JLab, the maximal $Q^2\sim6$~GeV$^2$ is too small
to suppress diagram $b$. 
Other higher-order diagrams might need to be considered as well.
At higher values $Q^2>10$~GeV$^2$, the process is completely dominated by 
the handbag diagrams with one hard propagator. The crossed, $u$-channel 
diagram $d$ of Fig.~\ref{Fig:diag} contributes on the 10\% level.  This 
should be contrasted with DIS, where this amplitude is 
very small and negligible.  In our calculations the skew diagrams $c$ and 
$e$ are of the same order as the regular handbag $a$ for all $Q^2$.  
Consequently, meson exchange mechanisms are likely to be 
important for the understanding of DVCS.

In the case of DIS, where cross sections are given by the imaginary part 
of the forward $\gamma^* N \to \gamma^* N$ amplitude, in the scaling 
region the $t$-channel meson exchange is relevant only at small $x_B$. 
 In the limit of large $s$ and low $t$, where $s=Q^2(1-x_B)/x_B$, 
 the amplitude is expected to be dominated by the right-most
 singularity in the complex-$l$ plane.  This leads to an amplitude 
 proportional to $\beta_n (Q^2) (Q/x_B m_N)^{\alpha_n(0)}$. 
 Due to the large photon virtuality, however, the 
 Regge residues $\beta_n(Q^2)$ are suppressed for high spin states 
 $\beta_n(Q^2) \sim (1/Q)^{\alpha_n(0)}$. This simply follows from the
 $t$-channel sub-process amplitudes, 
$\gamma^*(q)\gamma^*(q') \to n_l$, where $l = \alpha(m^2_n)$
\begin{equation}
\beta_n(Q) \sim A(\gamma^*\gamma^* \to n_l) \propto \int d^3 {\bf k} 
 P_l( \hat{\bf k}\cdot
 {\hat {\bf q}} ) {{\phi_n({\bf k})} \over {(k - (q+q')/2)^2}} 
  \propto \left({1\over Q}\right)^{l}.
\end{equation}
Thus, for DIS processes  the Regge limit only applies at low $x_B$, 
while at
 finite $x_B$ an alternative (parton) description becomes more
 efficient. In the case of DVCS, however, the presence of the real photon
 in the exit channel invalidates the above argument, since the residue
 functions, $\beta_n(Q) \sim A(\gamma^*\gamma \to n_l) \propto 
1/Q^2$ are not suppressed for high spin states~\cite{CZ}. In other 
words, the presence of 
 the soft $q \to N q$ fragmentation described by the bottom-right part of
 diagrams $c$ and $e$ masks the simple interpretation of the DVCS
 process in terms of partons originating from the nucleon alone.

We have shown that the perpendicular components of four vectors 
significantly alters the angular distribution of DVCS, and makes it 
skew toward backward angles. This effect is further enhanced by the 
inclusion of the skew diagrams.

This work will be extended to include a full treatment of the proton 
spin.  This will allow us to calculate the interference of DVCS with 
the Bethe-Heitler process and hence to calculate spin asymmetries, 
which are needed for a thorough comparison with data.  
With this extended model, we will again test the validity of commonly 
applied approximations, and we will evaluate the effectiveness of 
various proton wave functions.  We also intend to calculate 
meson photoproduction, with special
consideration of the meson poles that appear in the skew diagrams.

In conclusion, our calculations show that higher twist effects should 
be important in DVCS processes particularly at the relatively  
low energies currently accessible at JLab.  In this kinematic region, 
higher twist amplitudes need to be carefully studied and
calculated.  In addition, a substantial part of the intuition we 
bring from DIS processes may no longer be valid in evaluating 
DVCS processes. 

\begin{acknowledgments}
One of the authors (A.P.S.) would like to thank M. Burkardt for  
valuable discussions.  This work was supported in part by NSF grant 
NSF-PHY0070368 and DOE grant DE-FG02-87ER40365.
\end{acknowledgments}

\end{document}